\documentclass[aps,prd,showpacs,preprintnumbers,groupedaddress]{revtex4}
\usepackage{amsmath}
\usepackage{graphicx}
\usepackage{bm}
\begin{document}
\preprint{YITP-SB-03-57}
\preprint{BNL-HET-03/27}
\preprint{hep-ph/0311199}
\title{Scalar and Pseudoscalar Higgs Boson Plus One Jet Production \\
       at the LHC and Tevatron}
\author{B. Field}
\email[]{bfield@ic.sunysb.edu}
\affiliation{C.N. Yang Institute for Theoretical Physics, 
             Stony Brook University,
             Stony Brook, New York 11794-3840, USA}
\affiliation{Department of Physics, Brookhaven National Laboratory,
             Upton, New York 11973, USA}
\author{S. Dawson}
\email[]{dawson@bnl.gov}
\affiliation{Department of Physics, Brookhaven National Laboratory,
             Upton, New York 11973, USA}
\author{J. Smith}
\email[]{smith@insti.physics.sunysb.edu}
\affiliation{C.N. Yang Institute for Theoretical Physics, 
             Stony Brook University,
             Stony Brook, New York 11794-3840, USA}
\date{November 14, 2003}

\begin{abstract}

The production of the Standard Model (SM) Higgs boson ($H$) 
plus one jet is compared with that of the lightest scalar Higgs boson
($h^0$) plus one jet and that of the pseudoscalar Higgs boson ($A^0$) 
plus one jet. The latter particles belong to the Minimal
Supersymmetric Model (MSSM). We include both top and bottom quark loops to
lowest order in QCD and investigate the limits of small quark mass and
infinite quark mass.  We give results for both the CERN Large Hadron Collider 
(LHC) and the Fermilab Tevatron. 

\end{abstract}

\pacs{13.85.-t, 14.80.Bn, 14.80.Cp}
\maketitle

\section{Introduction} 
\label{intro}

The Higgs boson is the cornerstone of electroweak symmetry breaking in the
Standard Model (SM).  Particle physicists around the world have made the
search for the Higgs boson the top priority in high energy experiments.
However, there are several different candidate models in the Higgs sector.
The Minimal Supersymmetric Standard Model (MSSM), which is a special case
of the Two Higgs Doublet Model (2HDM), is of particular theoretical
interest.

The Standard Model Higgs boson has been experimentally excluded by LEP
searches for $e^+e^-\rightarrow ZH$ if its mass is lighter than
approximately $114$~GeV\!/c$^2$\cite{lepfinal}. In the MSSM, the
particle spectrum includes five physical Higgs bosons; a light and a
heavy neutral scalar ($h^0,H^0$), two charged scalars ($H^\pm$), and a
CP-odd pseudoscalar ($A^0$).  The mass of the lightest scalar in the
MSSM is excluded from being lighter than $91$~GeV\!/c$^2$\cite{neutral},
while the mass of the pseudoscalar $M_{A^{0}}$ is experimentally
excluded from being lighter than approximately $92$~GeV\!/c$^2$. The
ratio between the vacuum expectation values (VEVs) of the two neutral
Higgs bosons of the MSSM is defined as $\tan\beta = v_2/v_1$. For
$m_{\text{top}}=174.3$~GeV\!/c$^2$, $0.5 < \tan\beta < 2.4$ has been
excluded by the LEP Higgs searches. A different value of the top quark
mass will lead to different exclusion bounds on $\tan\beta$.

The total cross-section for scalar Higgs production including massive 
quark loops has been calculated at next-to-leading order (NLO)
in perturbative QCD \cite{spira1a, graudenz, spira1b}. 
The corresponding calculation for Higgs production in the MSSM
can be found in Ref.~\cite{spira2a}. In the Heavy Quark Effective Theory
(HQET)\cite{nanopoulos, hqet2, hqet1}, the top quark mass is assumed to
be much heavier than the Higgs boson mass and all relevant energy
scales. Assuming the HQET total inclusive cross-sections 
have been calculated at NLO for scalar\cite{sally} and pseudoscalar
production\cite{schaffer, spira2b} and also at NNLO for 
scalar\cite{harkil1, harkil2, anast1, jack2} and for
pseudoscalar\cite{jack2, harkil3, anast2} production, 
see also\cite{catani1,catani2}. 
The use of the HQET significantly simplifies the computation
of higher order QCD effects and has been shown to accurately reproduce
the exact NLO rate at the LHC for $pp\rightarrow H$\cite{spira1a,
spira1b} for a Higgs mass less than $1$~TeV\!/c$^2$ if the LO massive
results are multiplied by the NLO K-factor obtained in the HQET. 

In this paper we concentrate on the Higgs plus one jet ($gg \rightarrow
g\Phi$, $qg \rightarrow q\Phi$, and $q\bar{q} \rightarrow g\Phi$)
production processes since they are important for the experimental
detection of the Higgs. Here $\Phi$ represents either the SM Higgs, $H$, 
or the MSSM scalars, $h^0$ and $H^0$, or the MSSM pseudoscalar $A^0$. 
The production of the SM Higgs plus one jet process has been calculated 
exactly at LO in \cite{ellis, baur} with the inclusion of heavy quark loops. 
The production rate in the MSSM for the
lightest scalar plus one jet was recently calculated in LO
including SUSY loop effects, which can be significant for light SUSY
squarks and gluinos\cite{hollik}.  The NLO QCD
corrections to the Higgs plus one jet process have only been computed in
the HQET, since the full virtual corrections would require the
evaluation of massive two-loop integrals for a $2 \rightarrow 2$ reaction.
The differential cross-section for
the production of a scalar Higgs boson plus one jet in the HQET at NLO
has been calculated previously by \cite{catani1, catani2, higgscross,
jack, florian, glosser} and the integrated rate was shown to
increase substantially from the lowest order rate. The pseudoscalar case
has been presented in \cite{field2} and in \cite{kao}. 

We present the calculation of the Higgs plus one jet
process where we include both top and bottom quark loops with the full
quark mass dependence. This is done for the SM Higgs and for the
lightest scalar and pseudoscalar Higgs bosons of the MSSM.  
The contributions of loops with bottom quarks can be important for large
values of $\tan\beta$ in the MSSM. We also address the region of validity 
of the HQET predictions for these reactions.

In Section~\ref{limits}, the limit of the partonic matrix elements in
the HQET and in the small quark mass limit are explored. In
Section~\ref{full}, the Higgs plus jet matrix elements are given and our
computational techniques are described. Section~\ref{observables}
summarizes our notation for the hadronic differential cross-sections.
Section~\ref{results} contains numerical results for differential
cross-sections at the Tevatron and LHC, as well as integrated results
with cuts in transverse momemtum, $p_t$, and rapidity, $y$. Analytic
results for the matrix elements are given in two Appendices.

\begin{table*}
\begin{tabular}{c}
\begin{math}
|\mathcal{M}|^2 = | c_t^\Phi \mathcal{M}_t + 
                    c_b^\Phi \mathcal{M}_b |^2
\end{math}
\end{tabular} \\
\begin{tabular}{|c|l|l|} \hline
$\Phi$   & 
         & \\ \hline
$H$ &  \begin{tabular}{l}
            $c_t^H = \;\;\; 1$ \\ $c_b^H = \;\;\; 1$   
            \end{tabular}
         &  \begin{math}
            |\mathcal{M}|^2 = 
                    |\mathcal{M}_t|^2 
                  + |\mathcal{M}_b|^2 
                  + 2 \textrm{Re}
                          ( \mathcal{M}_t 
                            \mathcal{M}_b^\star )
            \end{math} \\
$h^0$    &  \begin{tabular}{l}
            $\displaystyle c_t^{h^{0}} = \;\;\; \cos\alpha / \sin\beta$ \\ 
            $\displaystyle c_b^{h^{0}} =      - \sin\alpha / \cos\beta$   
            \end{tabular}
         &  \begin{math}
            |\mathcal{M}|^2 = 
                    \displaystyle \frac{\cos^2 \! \alpha}{\sin^2 \! \beta} 
                    |\mathcal{M}_t|^2 +
                    \frac{\sin^2 \! \alpha}{\cos^2 \! \beta} 
                    |\mathcal{M}_b|^2 -
                    2 \frac{\sin\alpha\cos\alpha}{\sin\beta\cos\beta}
                    \textrm{Re} ( \mathcal{M}_t 
                                  \mathcal{M}_b^\star )
            \end{math} \\
$H^0$    &  \begin{tabular}{l}
            $\displaystyle c_t^{H^{0}} = \;\;\; \sin\alpha / \sin\beta$ \\ 
            $\displaystyle c_b^{H^{0}} = \;\;\; \cos\alpha / \cos\beta$   
            \end{tabular}
         &  \begin{math}
            |\mathcal{M}|^2 = 
                    \displaystyle \frac{\sin^2 \! \alpha}{\sin^2 \! \beta} 
                    |\mathcal{M}_t|^2 +
                    \frac{\cos^2 \! \alpha}{\cos^2 \! \beta} 
                    |\mathcal{M}_b|^2 +
                    2 \frac{\sin\alpha\cos\alpha}{\sin\beta\cos\beta}
                    \textrm{Re} ( \mathcal{M}_t 
                                  \mathcal{M}_b^\star )
            \end{math} \\
$A^0$    &  \begin{tabular}{l}
            $\displaystyle c_t^{A^{0}} = \;\;\; \cot\beta$ \\ 
            $\displaystyle c_b^{A^{0}} = \;\;\; \tan\beta$   
            \end{tabular}
         &  \begin{math}
            |\mathcal{M}|^2 = 
                     \displaystyle
                     \frac{1}{\tan^2 \! \beta} |\mathcal{M}_t|^2 +
                              \tan^2 \! \beta  |\mathcal{M}_b|^2 +
                     2 \textrm{Re} ( \mathcal{M}_t 
                                     \mathcal{M}_b^\star )
            \end{math} \\ \hline
\end{tabular}
\caption{Higgs-fermion couplings in the MSSM and the dependence of the
matrix element-squared on the couplings. $\mathcal{M}_t$ and
$\mathcal{M}_b$ represent contributions from top- and bottom- quark loops,
respectively.  The $\alpha$ parameter is the angle that diagonalizes the
CP-even Higgs squared-mass matrix.}
\label{table}
\end{table*}

\section{Partonic Processes - Heavy Quark Effective Theory}
\label{limits}

\begin{figure}
  \begin{center}
    \begin{tabular}{cc}
      \resizebox{60mm}{!}{\includegraphics{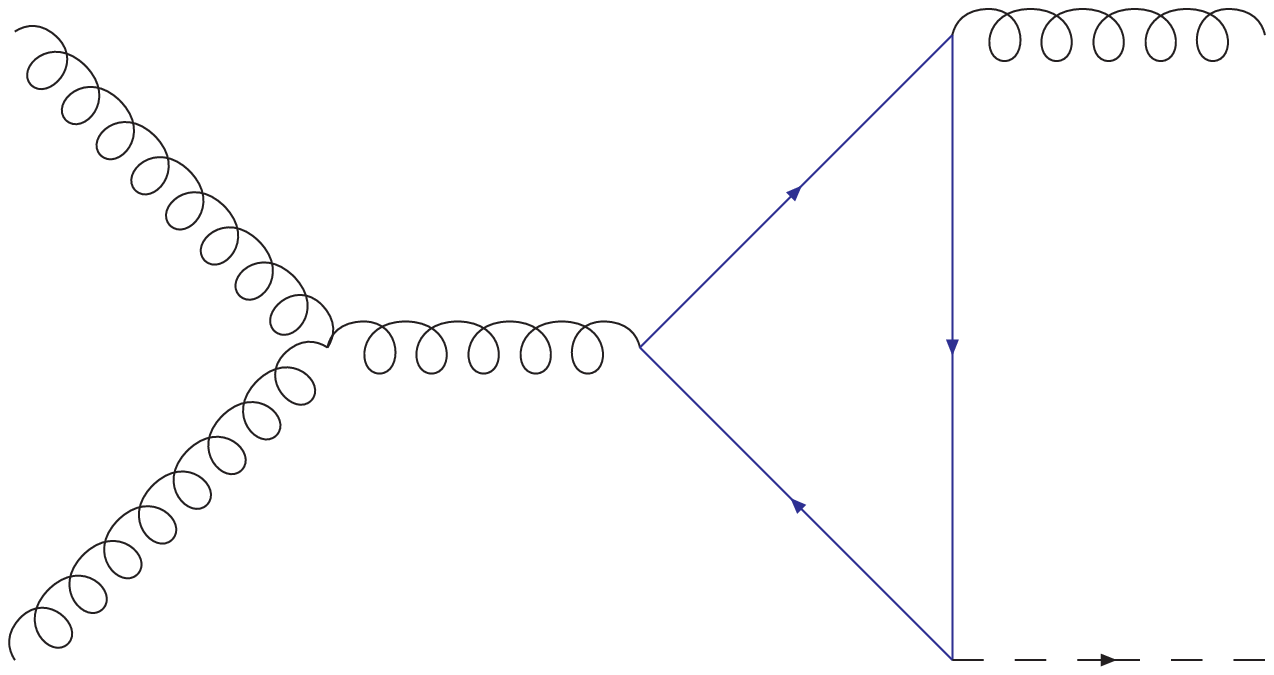}} &
      \resizebox{55mm}{!}{\includegraphics{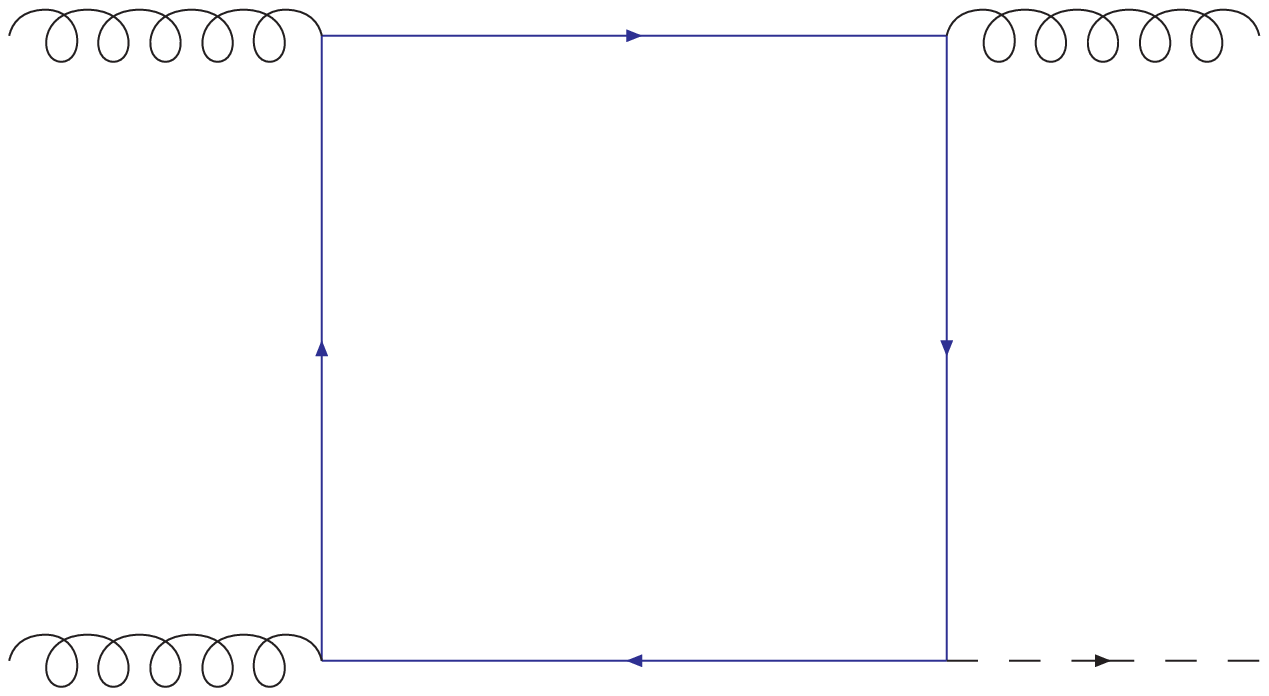}} \\
      (a) & (b) \\
          &     \\
      \resizebox{60mm}{!}{\includegraphics{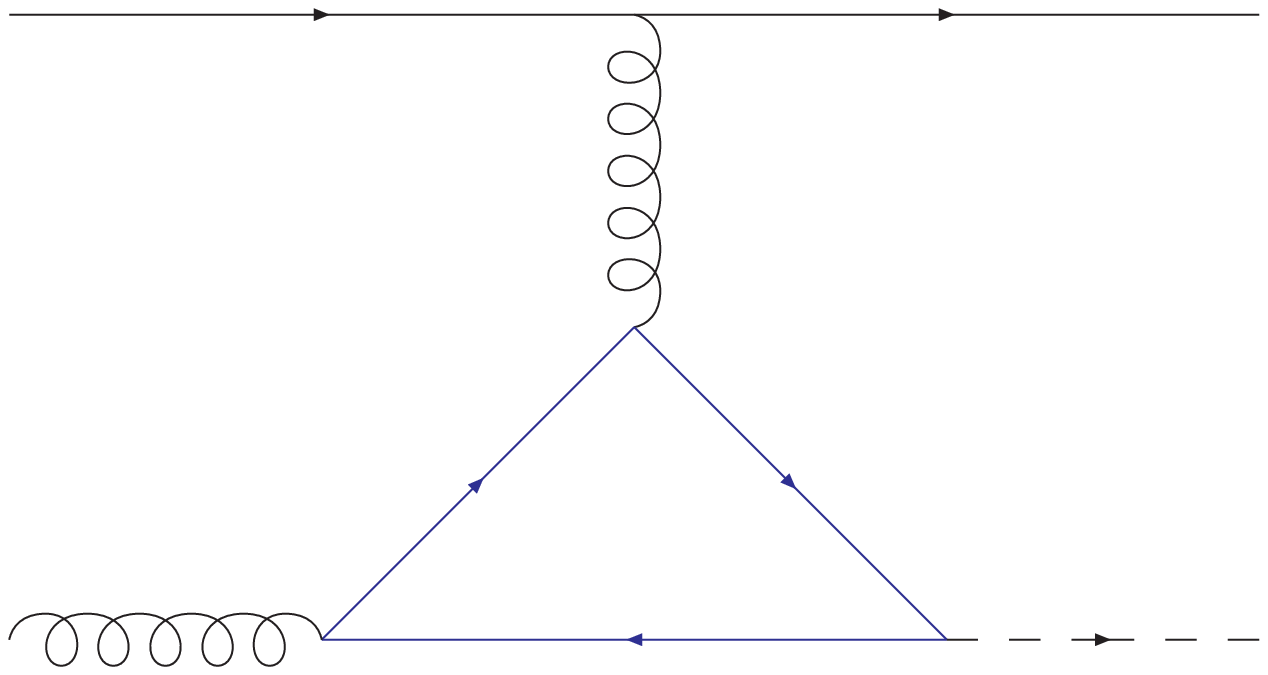}} &
      \resizebox{60mm}{!}{\includegraphics{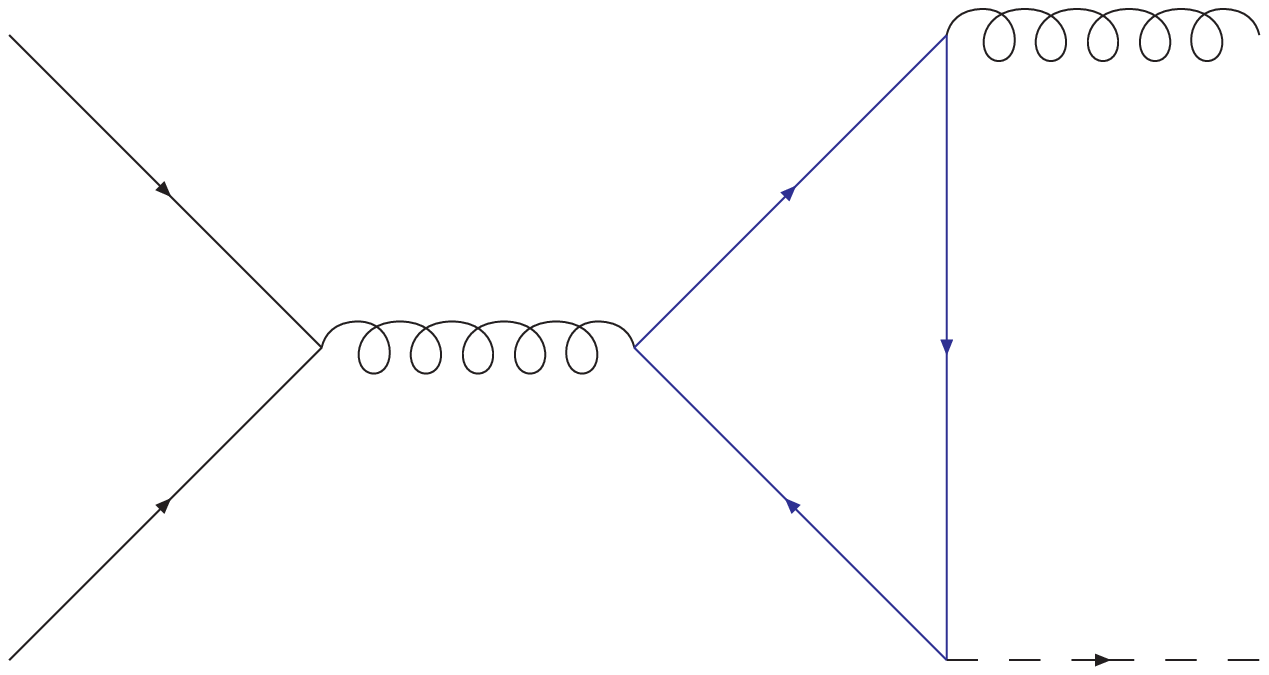}} \\
      (c) & (d) \\
    \end{tabular}
    \caption{Sample Higgs plus one jet diagrams. Figures 1a,b are the $gg$ 
             diagrams, 1c is the $qg$ channel, and 1d is the $q\bar{q}$ 
             channel. All quarks contribute to the loops. The crossed and
             charge conjugate diagrams are not
             shown. There are a total of $12$ $gg$ diagrams and $2$ for 
             each of the $qg$ and $q\bar{q}$ sub-processes. }
    \label{diags}
  \end{center}
\end{figure}

In the limit where the top quark mass is much heavier than all the
energy scales in the problem, only the top quark coupling to gluons is
numerically significant and this limit provides a good approximation to
Standard Model Higgs production matrix elements. The HQET limit for
scalar Higgs production has been extensively studied in the literature.
This limit is especially useful for deriving higher order QCD
corrections since the massive top quark loops that couple the Higgs
boson to gluons reduce to effective vertices. The Feynman rules
can be derived from an effective Lagrangian density\cite{spira1a,
spira1b, hqet1, hqet2, sally, harkil1},
\begin{equation}
\mathcal{L}_{\text{eff}}^H = -g_H\frac{H}{4v} {\mathcal C}_H(\alpha_s)
 \mathcal{O}_H, \quad 
 \mathcal{O}_H = G^a_{\mu\nu} G^{a,\mu\nu},
\label{leffH}
\end{equation}
where $g_H=1$ in the Standard Model and $v=246$~GeV.  ${\mathcal O}_H$
generates vertices which couple the Higgs boson to two, three, and four
gluons. In the large $m_\text{top}$ limit, the 
coefficient
${\mathcal C}_H$ can be evaluated as a power series in
$\alpha_s$\cite{spira1a, spira1b, hqet2, hqet1, hqet_h1, hqet_h2,
hqet_h3}
\begin{equation}
\mathcal{C}_H (\alpha_s^{(5)}(\mu_r^2)) =  
-\frac{\alpha_s^{(5)}(\mu_r^2)}{3\pi}\biggl[1+\frac{11\alpha_s^{(5)}(
\mu_r^2)}{4\pi} +\cdots\biggr],
\end{equation} 
where $\alpha_s^{(5)}(\mu_r^2)$ is evaluated at the scale $\mu_r$ in a $5$
flavor scheme.

For comparison, we consider a pseudoscalar Higgs boson with a coupling
to fermions given by,
\begin{equation}
\mathcal{L}_{\text{eff}}^{A^0} = -ig_A \frac{{A^0}}{v} m_i
{\overline \psi}_i \gamma_5 \psi_i .
\label{pseudodef}
\end{equation}
In the large $m_\text{top}$ limit, the interactions of the pseudoscalar 
with gluons can be found from the effective Lagrangian \cite{harkil2, 
russel2, footnote, hqet_a1} 
\begin{equation}
\mathcal{L}_{\text{eff}}^{A^0} = -g_A \frac{{A^0}}{v}
\bigl( C_{A_{1}}(\alpha_s) \mathcal{O}_1 
+ C_{A_{2}}(\alpha_s) \mathcal{O}_2 \bigr), \quad  
\mathcal{O}_1 = \epsilon_{\mu\nu\lambda\sigma}
G^{\mu\nu}_a G^{\lambda\sigma}_a, \quad
\mathcal{O}_2 = \partial^\mu 
\sum^{n_f}_{i=0} \bar{q}_i \gamma_\mu\gamma_5 q_i,
\label{leffa}
\end{equation}
where $G^{\mu\nu}_a$ is the gluon field strength tensor.
The process independent coefficient functions are
\begin{equation}
C_{A_{1}} = - \frac{\alpha_s(\mu_r^2)}{16\pi}, \qquad
C_{A_{2}} = \; {\cal O}(\alpha_s^2).
\end{equation}
We consider $g_A=1$ and the examine the differences between 
differential cross-sections for the production of a SM scalar
Higgs boson and a pseudoscalar Higgs boson with the couplings of Eq.
\ref{pseudodef}, when the bosons are produced in association with a jet.

It is also of interest to compare the production rates for a Higgs boson
plus a jet in the MSSM. The effective Lagrangians in this case are found
by making the replacements in Eqs.~\ref{leffH} and \ref{leffa},
\begin{align}
g_H\rightarrow & c_t^{h^{0}}\nonumber \\
g_A\rightarrow & c_t^{A^{0}},
\label{coupdef}
\end{align}
where $c_t^{h^{0}}$ and $c_t^{A^{0}}$ are given in Table I. (We neglect
contributions from SUSY particles such as the bottom squarks and gluinos,
and therefore assume that the SUSY particle masses are much larger than
$m_\text{top}$ and $m_\Phi$. These genuine SUSY contributions can be
important for light squark and gluino masses\cite{hollik}.)  When the
bottom quark becomes important, the HQET breaks down as a reliable
calculational tool. This occurs in the MSSM when $\tan\beta$ becomes large
and the bottom quark couplings are enhanced.

\begin{figure}
  \begin{center}
    \begin{tabular}{c}
      \includegraphics[scale=0.77]{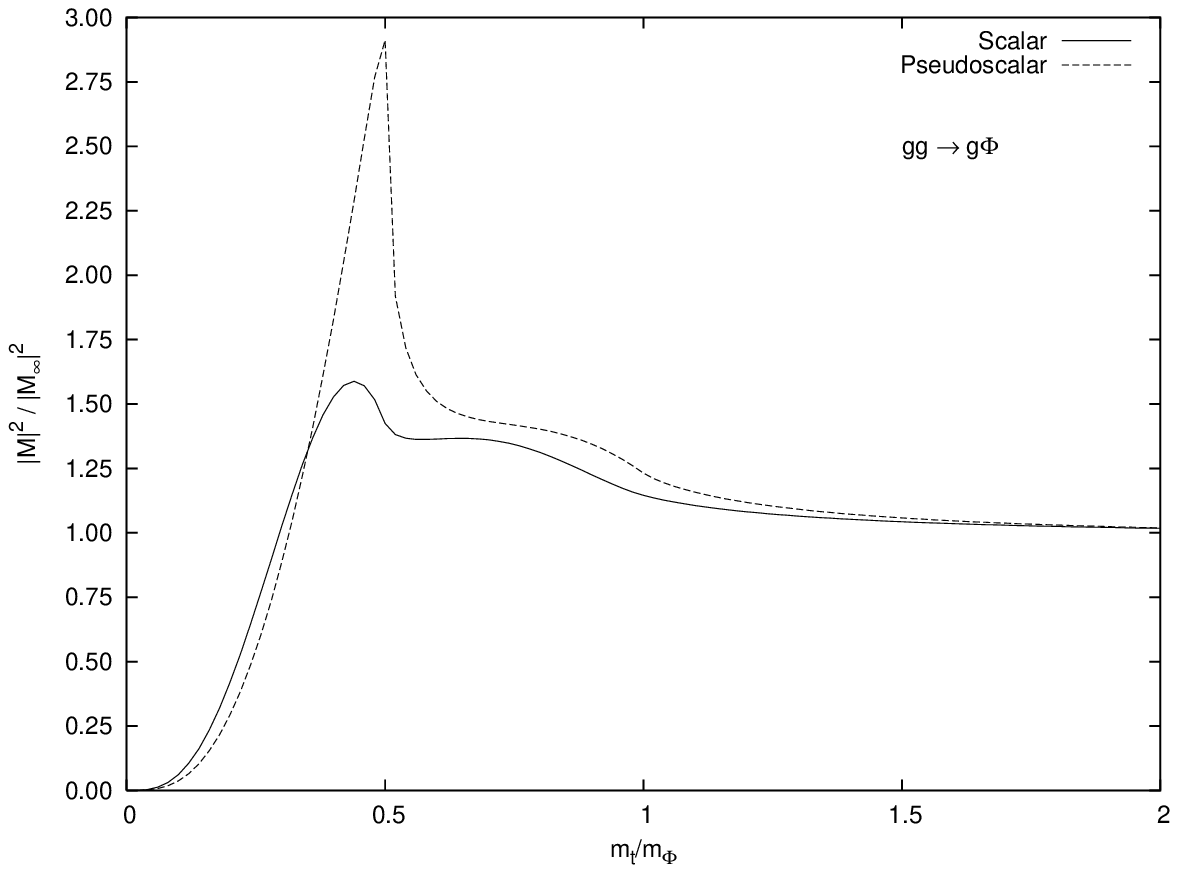} \\
      \includegraphics[scale=0.77]{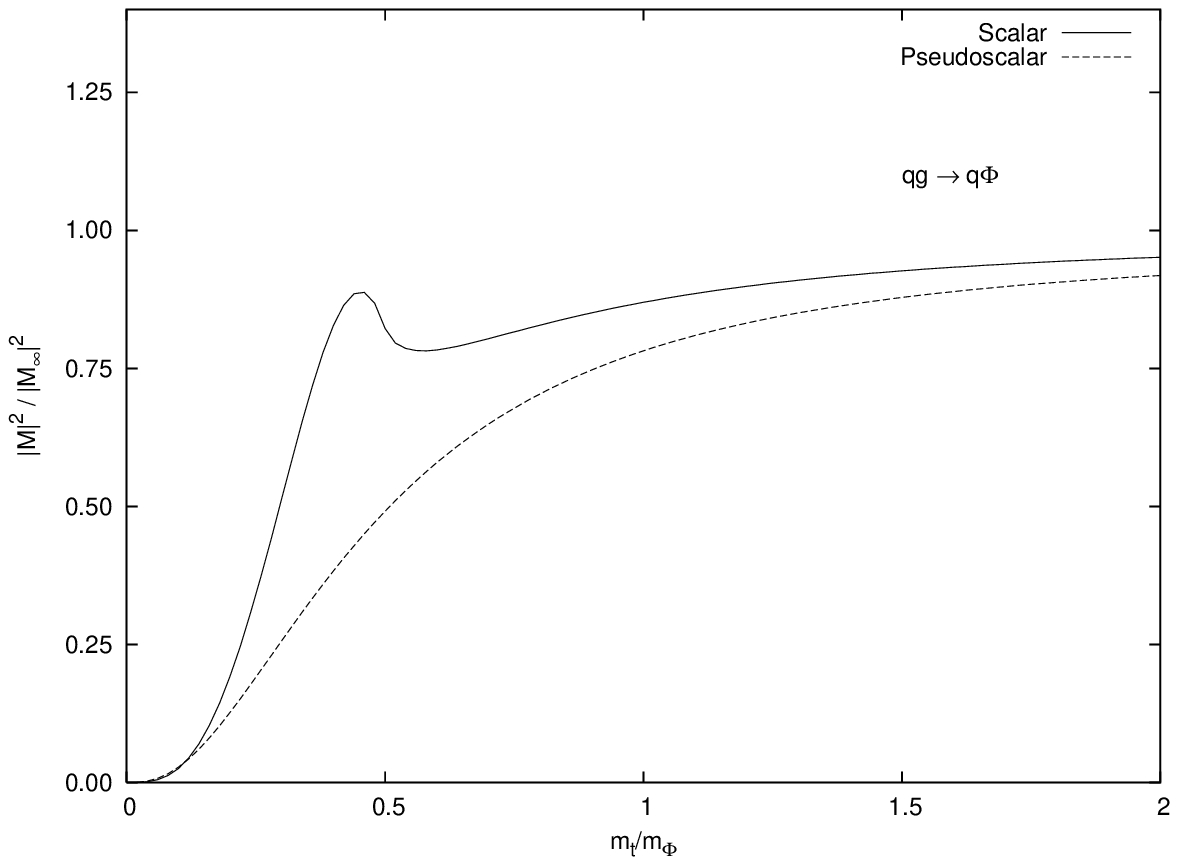} \\
      \includegraphics[scale=0.77]{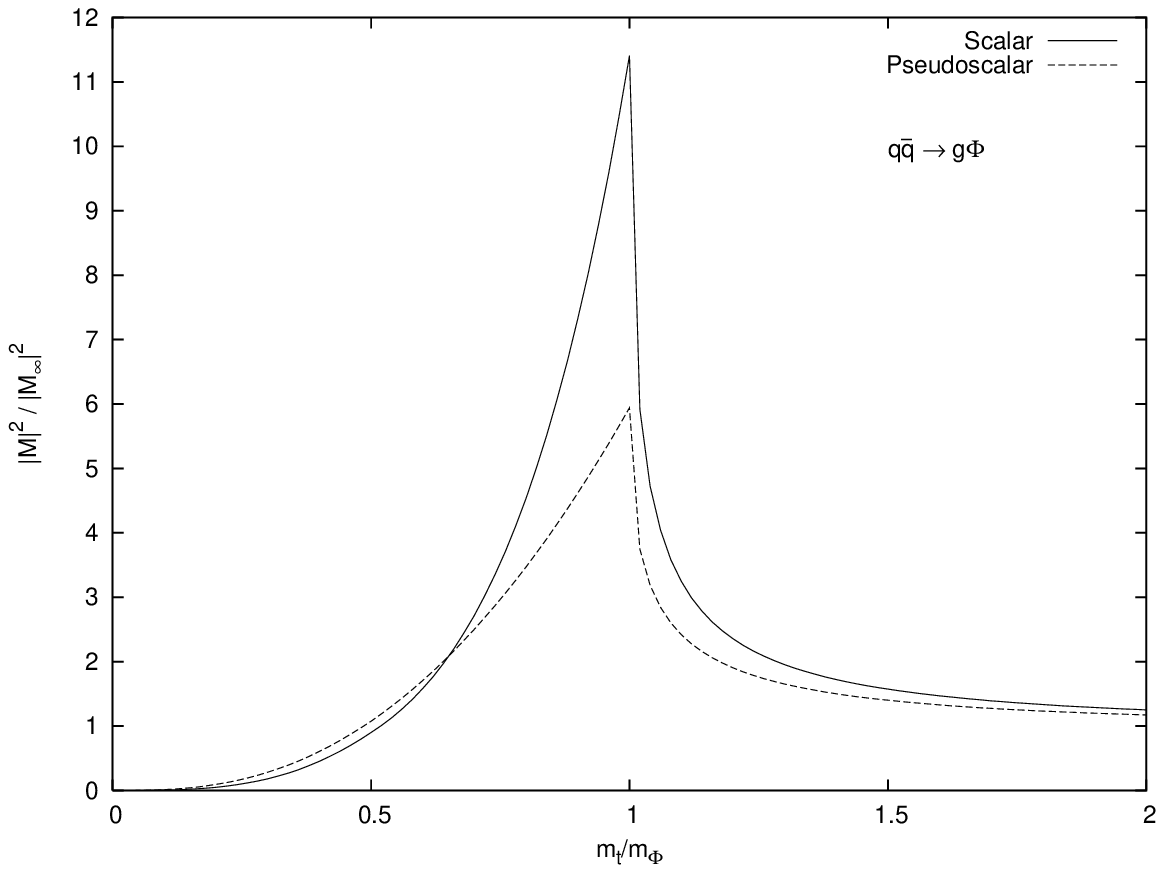} 
    \end{tabular}

\caption{The squared matrix elements, $|\mathcal M|^2$, evaluated at
         $\hat{s} = 4m_\Phi^2$ and $\hat{u}=\hat{t}$ for the three 
         different channels, ($gg$, $qg$, $q\bar{q}$), normalized to the 
         squared matrix elements in the HQET for  scalar and 
         pseudoscalar (with $g_A = 1$) Higgs plus jet production.
         We include only the top quark loops. 
         The solid line is the scalar, whereas the dashed 
         line is the pseudoscalar.} 
    \label{ratios}
  \end{center}
\end{figure}

\section{Partonic Processes - Full Theory}
\label{full}

There are three channels associated with Higgs plus one jet production:
gluon fusion, quark-gluon scattering, and quark-antiquark annihilation.  
Representative Feynman diagrams are shown in Fig.~\ref{diags}. At the LHC
where $\sqrt{S}=14$~TeV the gluon fusion and quark-gluon channels are the most
important, with the quark-antiquark channel adding a negligible amount to
the process. However all three channels are important at the Tevatron
where $\sqrt{S} =1.96$~TeV.

The calculation of the matrix elements was carried out in both
$n=4-2\epsilon$ dimensions and $4$-dimensions.  The $\gamma_5$ in the
pseudoscalar calculation was treated using the Akyeampong-Delbourgo
prescription\cite{ad1,ad2,ad3} for the $\gamma_5$-matrix. In this scheme
the $\gamma_5$ is exchanged for a Levi-Civita tensor contracted with four
$\gamma$-matrices.  After the trace, the tensor loop integrals were
reduced to scalar integrals using the usual Passarino-Veltman\cite{loops}
reduction techniques.

\subsection{Gluon fusion ($gg \rightarrow g\Phi$)}
The gluon fusion channel is the most important channel at the LHC. The 
momentum distribution in this process is assigned with all momentum 
incoming,
\begin{equation}
g( p_1^{\mu_1,a_1}) + g(p_2^{\mu_2,a_2}) \rightarrow 
g(-p_3^{\mu_3,a_3}) + \Phi(-p_5),
\end{equation}
where $\mu_i$ are Lorentz indices and $a_i$ are color indices. The 
Mandelstam variables used in the partonic system are
\begin{equation}
      \hat{s} = (p_1+p_2)^2,
\quad \hat{t} = (p_1+p_5)^2,
\quad \hat{u} = (p_2+p_5)^2,
\quad Q^2     = m^2_{\Phi}.
\end{equation}
The matrix elements, including the gluon polarization vectors, can be 
written
\begin{equation}
\mathcal{M}^{gg} = \epsilon^{\mu_1}(p_1)
                   \epsilon^{\mu_2}(p_2)
                   \epsilon^{\mu_3}(p_3)
                   \mathcal{M}^{gg}_{\mu_1 \mu_2 \mu_3}.
\end{equation}
The Ward-Takahashi identities let us check the gauge invariance of the 
sub-process. In the gluon fusion case, these can be written as
\begin{equation}
p_1^{\mu_1} \epsilon^{\mu_2}(p_2)\epsilon^{\mu_3}(p_3)
\mathcal{M}^{gg}_{\mu_1 \mu_2 \mu_3} =
\epsilon^{\mu_1}(p_1)p_2^{\mu_2}\epsilon^{\mu_3}(p_3)
\mathcal{M}^{gg}_{\mu_1 \mu_2 \mu_3} =
\epsilon^{\mu_1}(p_1)\epsilon^{\mu_2}(p_2)p_3^{\mu_3}
\mathcal{M}^{gg}_{\mu_1 \mu_2 \mu_3} = 0,
\end{equation}
giving us a strong check on the algebraic results. Analytic results  for 
the matrix element squared for $gg\rightarrow g A^0$ are 
given in Appendix A, see also Appendix C in \cite{spira1b}, 
while those for $gg\rightarrow gH$ can be found in 
Refs.~\cite{baur,ellis}.

\subsection{Quark-antiquark annihilation ($q\bar{q} \rightarrow g\Phi$)}
For this sub-process, the momentum, color, and Lorentz structure was 
assigned as follows
\begin{equation}
q(p_1) + \bar{q}(p_2) \rightarrow g(-p_3^{\mu_3,a_3}) + \Phi(-p_5).
\end{equation}
The matrix elements satisfy
\begin{equation}
\mathcal{M}^{q\bar{q}} = \epsilon^{\mu_3}(p_3)
                   \mathcal{M}^{q\bar{q}}_{\mu_3},
\qquad
p_3^{\mu_3} \mathcal{M}^{q\bar{q}}_{\mu_3} = 0.
\end{equation}
Analytic results for $q {\overline q} \rightarrow g A^0$ are given in
Appendix B, see also Appendix C in \cite{spira1b},
while those for $q {\overline q} \rightarrow g H$ can be found in
Refs. \cite{baur,ellis}. The results for quark-gluon scattering can be
found by crossing.

\subsection{HQET Matrix Elements}
The $4$-dimensional color-spin averaged matrix elements for Higgs boson
plus one jet production in the $m_\text{top}\rightarrow \infty$ limit are
presented here for completeness. These matrix elements obey the same
crossing relations as the full matrix elements,
\begin{equation}
 |\mathcal{M}(\hat{s},\hat{t},\hat{u})|^2_{qg\rightarrow q \Phi}
 = -|\mathcal{M}(\hat{u},\hat{t},\hat{s})|^2_{q\bar{q}\rightarrow g \Phi}.
\end{equation}
The matrix elements in the large $m_\text{top}$ HQET 
limit can be written\cite{spira1b,baur,ellis,kao},
\begin{align}
\overline{\sum} |\mathcal M |^2_{gg\rightarrow g\Phi}
 &= A_\Phi \frac{N_c}{4(N_c^2-1)} \;
                \frac{\hat{s}^4+\hat{t}^4+\hat{u}^4+Q^8}
                     {\hat{s}\hat{t}\hat{u}} \\
\overline{\sum} |\mathcal M |^2_{qg\rightarrow q\Phi}
 &= -A_\Phi\frac{1}{8 N_c} \;
                \frac{\hat{s}^2+\hat{t}^2}{\hat{u}} \\
\overline{\sum} |\mathcal M |^2_{q\bar{q}\rightarrow g\Phi}
 &= A_\Phi \frac{(N_c^2-1)}{8N_c^2} \;
                \frac{\hat{u}^2+\hat{t}^2}{\hat{s}},
\end{align}
where,
\begin{align}
A_H=& \biggl( \frac{\alpha_s}{3 \pi v}\biggr)^2 (4 \pi \alpha_s) g_H^2
\nonumber \\
A_A=& \biggl( \frac{\alpha_s}{2 \pi v}\biggr)^2 (4 \pi \alpha_s) g_A^2
\end{align}
and $g_H=1$ for the SM and $g_\Phi$ is given in Eq. \ref{coupdef} for the
MSSM. The bar implies a sum and average over colors and spins. The
exact matrix elements squared as compared with the HQET matrix elements
are shown in Fig.~\ref{ratios} for both the SM scalar, which are in excellent
agreement with the plots in \cite{baur}, and for a
pseudoscalar with $g_A=1$. In this plot, the mass of the Higgs was set to
$m_\Phi=100$~GeV\!/c$^2$ and the mass of the top quark was varied. In
these plots, two thresholds can be observed. Each threshold occurs when an
imaginary part of the matrix elements turns on or off. If we examine 
Eq.~\ref{qqgA} for $q\bar{q} \rightarrow gA^0$ we clearly see that the 
imaginary part contains the difference of two step functions
\begin{equation}
\theta(     \hat{s} - 4 m_\text{top}^2 ) - 
\theta( M^2_{A^{0}} - 4 m_\text{top}^2 ),
\end{equation}
so the first threshold occurs at $2m_\text{top} = M_{A^{0}}$ and the
second at $2m_\text{top} = \sqrt{\hat{s}}$. Since we choose
$\hat{s}=4M_{A^{0}}^2$ for the plot this implies that these thresholds
occur at $m_\text{top}/m_\Phi=0.5$ and $1$ respectively. The imaginary
part is finite between these cusps. Similar phenomena occur in the other
reactions. However when the squared matrix elements
contain several terms the onset of the imaginary parts is 
not always visible. The reactions $qg
\rightarrow q\Phi$ do not have $\hat{s}$ channels so they only have 
cusps at $m_\text{top}/m_\Phi=0.5$. Finally the $gg \rightarrow g\Phi$
channels show both cusps. Note that the reason the cusps do not appear
exactly at $0.5$ and $1$ is due to our choice of points 
in $m_\text{top}/m_\Phi$. 

These ratios show that when the heavy quark becomes heavier than
$m_\text{top} \sim \frac{1}{2} m_\Phi$ the HQET is a reasonable
approximation to the matrix elements with a top loop only. In the MSSM,
however, the usefulness of the HQET is limited to small values of
$\tan\beta$ where the bottom quark contribution can be neglected.

\subsection{Small Quark Mass Limit}

When the quark mass in the loop is much smaller than the Higgs mass and
the energy scale, the small quark mass limit $m_f \rightarrow 0$ is
relevant.  This is the case for the bottom quark contribution in the large
$\tan\beta$ limit of the MSSM. The matrix elements in this limit behave as
\begin{equation}
|\mathcal M|^2 \sim m_{f}^4 \log^4(m_f^2/\mu^2),
\end{equation}
where $\mu >\!\!> m_f$. Exact expressions in the small quark mass limit 
are given in Appendix B.

\section{Observables}
\label{observables}

Generically, we can write a $2 \rightarrow 2$ differential observable as
\begin{equation}
\hat{s}^2 \frac{d^2 \! \hat{\sigma}}{d\hat{t}\;d\hat{u}} = 
\frac{1}{16\pi} \; \overline{\sum} |\mathcal{M}|^2,
\end{equation}
where the bar implies a sum and average over colors and spins. To 
relate the hadronic differential distributions to the partonic differential 
distributions we need to perform a convolution with the parton 
distribution functions. 

The hadronic process can be written as
\begin{equation}
H_1(P_1) + H_2(P_2) \rightarrow j(-p_3) + \Phi(-p_5)
\end{equation}
where the $j$ represents the gluon or the quark jet in the sub-process of 
interest. In the hadronic system, we can write
\begin{equation}
S=(P_1+P_2)^2, \quad
T=(P_1+p_5)^2, \quad
U=(P_2+p_5)^2.
\end{equation}
This translates into the partonic system (with momentum fractions $x_1$ 
and $x_2$) as
\begin{align}
& p_1 = x_1 P_1, \quad
  p_2 = x_2 P_2, \\
& \hat{s} = x_1 x_2 S, \quad
  \hat{t} = x_1 ( T-Q^2 ) + Q^2, \quad
  \hat{u} = x_2 ( U-Q^2 ) + Q^2 \\
& x_{1,\text{min}} = \frac{-U}{S+T-Q^2}, \quad
  x_{2,\text{min}} = \frac{-x_1 (T-Q^2)-Q^2}{x_1 S + U - Q^2},
\end{align}
where $Q^2=m_\Phi^2$.
The hadronic variables can be written in terms of the transverse momentum 
and rapidity
\begin{align}
T &= Q^2 - \sqrt{S}\sqrt{p_t^2+Q^2}\cosh y
         + \sqrt{S}\sqrt{p_t^2+Q^2}\sinh y \\
U &= Q^2 - \sqrt{S}\sqrt{p_t^2+Q^2}\cosh y
         - \sqrt{S}\sqrt{p_t^2+Q^2}\sinh y.
\end{align}
The hadronic differential cross-section is,
\begin{align}
S^2\frac{d^2 \! \sigma^{H_{1} H_{2}}}{dT \; dU} = 
\sum_{ab} \int^1_{x_{1,\text{min}}} \frac{dx_1}{x_1} 
          \int^1_{x_{2,\text{min}}} \frac{dx_2}{x_2} \;
          f^{H_{1}}_a(x_1,\mu_f^2) f^{H_{2}}_b(x_2,\mu_f^2 ) \;
          \hat{s}^2 \frac{d^2 \! \hat{\sigma}_{ab}}{d\hat{t} \; d\hat{u}}.
\end{align}
Upon further integration we obtain the single differential $p_t$ and
rapidity distributions with the kinematic limits,
\begin{align}
& p_{t,\text{max}} = \frac{1}{2} \frac{S-Q^2}{\sqrt{S}}, \quad
  y_\text{max} = \frac{1}{2}\ln \biggl( \frac{1+S_Q}{1-S_Q} \biggr), \\
& \text{where} \quad S_Q = \sqrt{ 1-\frac{4S(p_t^2+Q^2)}{(S+Q^2)^2} }.
\end{align}

\section{Numerical Results}
\label{results}

We present our calculations for the CERN LHC with $\sqrt{S} = 14$~TeV and
the Fermilab Tevatron with $\sqrt{S} = 1.96$~TeV. We use the CTEQ6.1L parton
distribution functions\cite{cteq} with $\Lambda_5^{\text{LO}} =
165.2$~MeV and a one loop running coupling constant with $\alpha_s(M_Z)
= 0.1298$. For the differential distributions, the full kinematic
rapidity and $p_t$ are used and the factorization and renormalization
scales are set equal to,
\begin{equation}
\label{murf}
\mu_r = \mu_f = \sqrt{ Q^2 + p_t^2 }.
\end{equation}
We use pole masses with $m_\text{top}=174.3$~GeV\!/c$^2$ and
$m_\text{bot}=4.5$~GeV\!/c$^2$. For the integrated cross-section we
require the $p_t$ of the Higgs and the jet to satisfy $p_{t,\text{min}}
> 30$~GeV\!/c in the rapidity region $|y|<2.5$ and replace $p_t$ 
by $p_{t,\text{min}}$ in Eq.~\ref{murf} for the renormalization and 
factorization scales.

\subsection{Standard Model}

\begin{figure}
\includegraphics{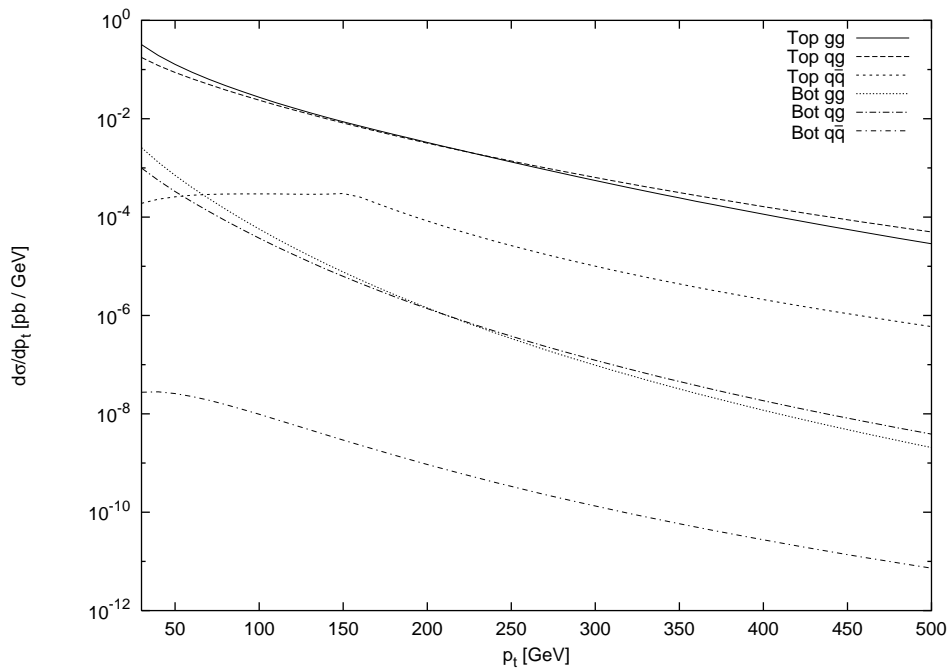}
\caption{Transverse momentum distributions for the SM 
Higgs boson plus one jet
production at the LHC with $M_{H}=120$~GeV\!/c$^2$ for the different 
channels.  The curves labeled `Top' (`Bot') include ${\it only}$ the top
(bottom) quark loops.}
\label{sm_pt}
\end{figure}

\begin{figure}
\includegraphics{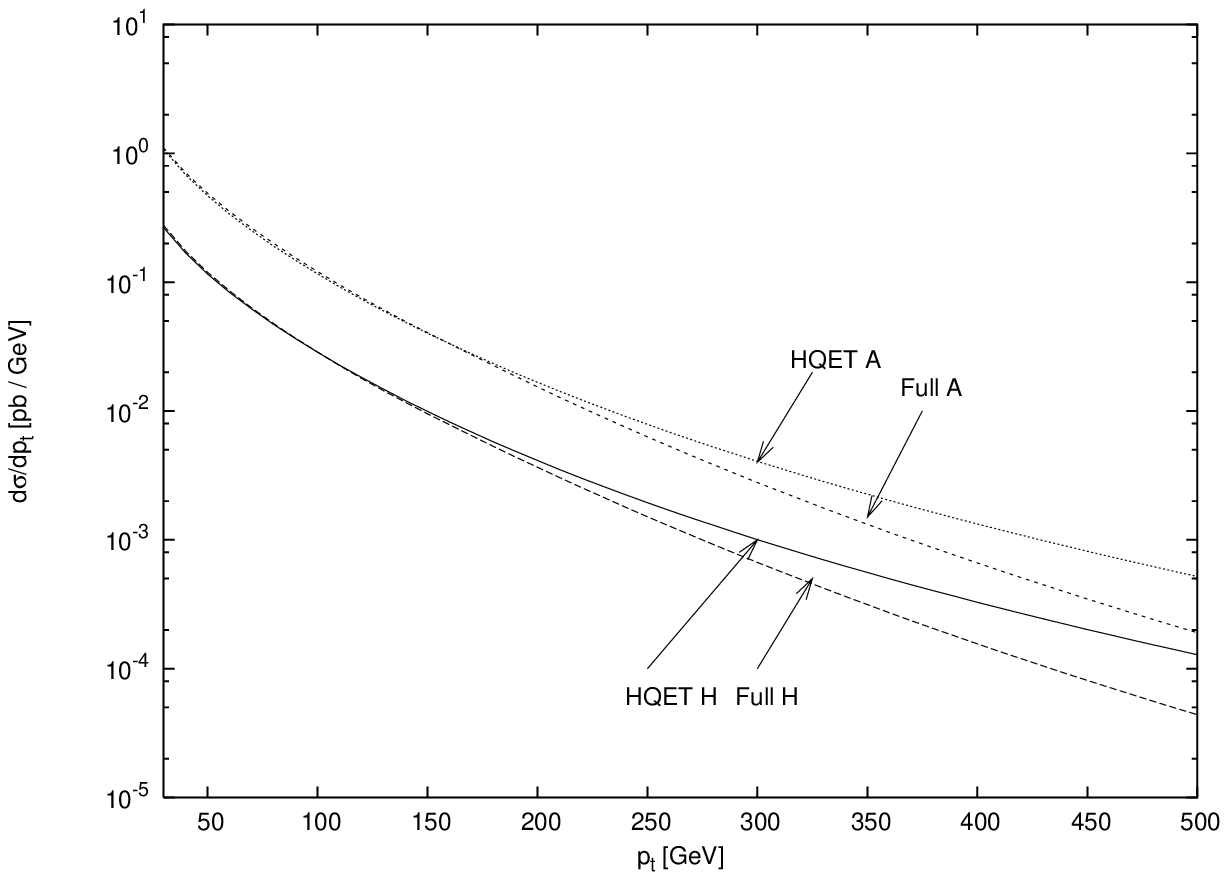}
\caption{Transverse momentum distributions for the SM Higgs ($H$) plus
one jet and for a pseudoscalar ($A^0$) plus one jet in the full theory
with only the top-quark loops included and in the HQET at the LHC for
$M_{\Phi}=120$~GeV\!/c$^2$. We assume $g_A=1$.}
\label{full_hqet}
\end{figure}

The transverse momentum distributions of the SM Higgs boson for all the
separate channels are shown in Fig.~\ref{sm_pt} for the LHC.  For a SM
Higgs boson with $M_{H} = 120$~GeV\!/c$^2$, the cross-section for Higgs
plus one jet is approximately $12.3$~pb when both the top and bottom
quarks are included in the calculation. Although the bottom quark
contribution alone is only $0.05$~pb, the top-bottom interference lowers
the cross-section by approximately $8.25$\% from $13.4$~pb when
only the top quark is included, see \cite{sally2}. 
This lowering of the cross-section may be visible at the LHC. 
As shown in Fig.~\ref{full_hqet}, the full
theory and the HQET agree very well at small to moderate $p_{t}$ for
both the scalar\cite{ellis, glosser} and the pseudoscalar differential 
distributions.

\begin{figure}
  \begin{center}
    \begin{tabular}{c}
      \includegraphics{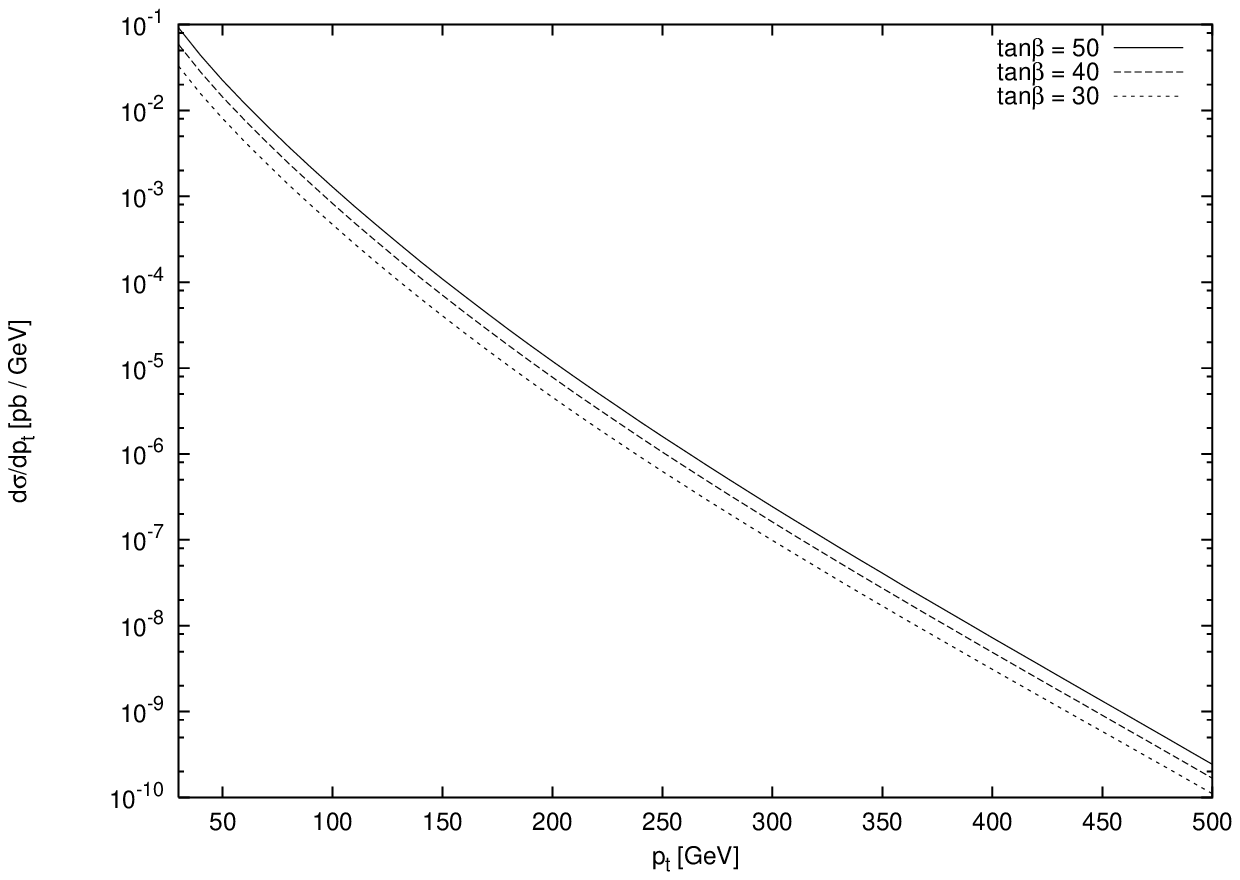} \\
      \includegraphics{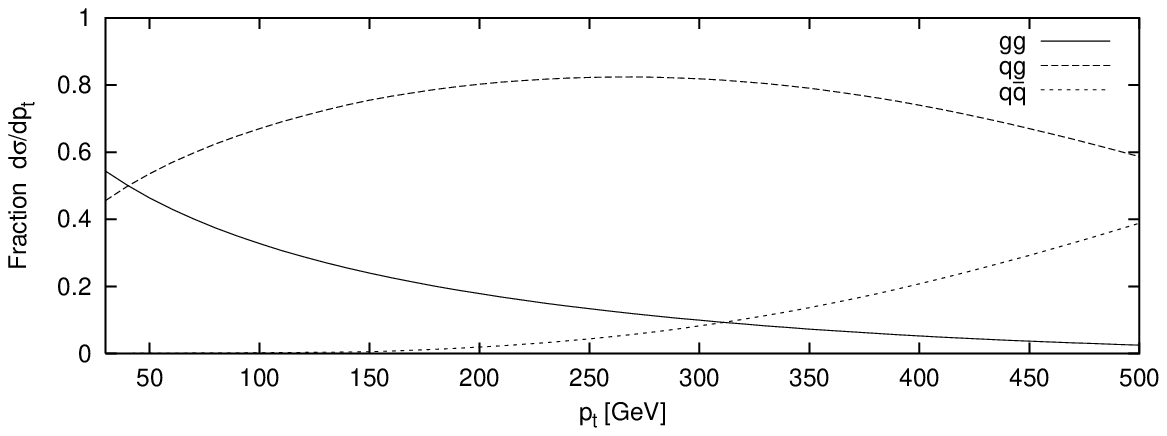}
    \end{tabular}
    \caption{The transverse momentum distributions for the MSSM
             pseudoscalar Higgs boson for $\tan\beta=30,40,50$ and 
             $M_{A^{0}} = 120$~GeV\!/c$^2$ at the Tevatron  including the 
             top and bottom quark loops. The top, middle, and bottom lines 
             in the top graph are the curves for $\tan\beta = 50,40,30$ 
             respectively. Below 
             is the fraction of the process that comes from each of the 
             different channels. The curves at $p_t = 250$~GeV\!/c from 
             top to bottom are the $qg$, $gg$, and $q\bar{q}$ channels 
             respectively.}
    \label{tevatron}
  \end{center}
\end{figure}

\subsection{Minimal Supersymmetric Standard Model}

The MSSM is a special case of the 2HDM. In the MSSM, the up- and
down-type quarks become massive from different Higgs doublets and the
ratio of the two VEVs is parameterized by $\tan\beta=v_2/v_1$. As shown
in Table~\ref{table}, up- and down-type quarks couple differently to the
Higgs bosons of the MSSM. The $\alpha$ parameter is the angle that is
introduced to diagonalize the mass eigenstates of the CP-even Higgs
squared-mass matrix to obtain the physical states. The program
HDECAY\cite{hdecay} was used to determine the mass of the lightest
scalar and the $\alpha$ mixing parameter once the values of 
$M_{A^{0}}$ and $\tan\beta$ were chosen. The SUSY Higgs mixing parameter
was set to $\mu = 300$~GeV\!/c$^2$, the gluino mass to $\mu_2 =
200$~GeV\!/c$^2$, all the SUSY breaking masses to $1$~TeV\!/c$^2$, and
the soft breaking term to $1.5$~TeV\!/c$^2$.

At the Tevatron, there is a very small signal for the SM Higgs boson.
The cross-section for a SM Higgs boson plus one jet with $M_{H} =
120$~GeV\!/c$^2$ at lowest order in QCD is approximately $0.1$~pb. For
$\tan\beta \sim 30$ the cross-section for a $120$~GeV\!/c$^2$
pseudoscalar Higgs in the MSSM is about twice as large as for a
$120$~GeV\!/c$^2$ SM Higgs at the Tevatron and continues to grow with
$\tan\beta$. The differential cross-section for pseudoscalar plus jet
production at the Tevatron is shown in Fig.~\ref{tevatron}. At the
Tevatron, the large $\tan\beta$ region is completely dominated by bottom
quark loops where the HQET is of little use.

\begin{figure}
\includegraphics{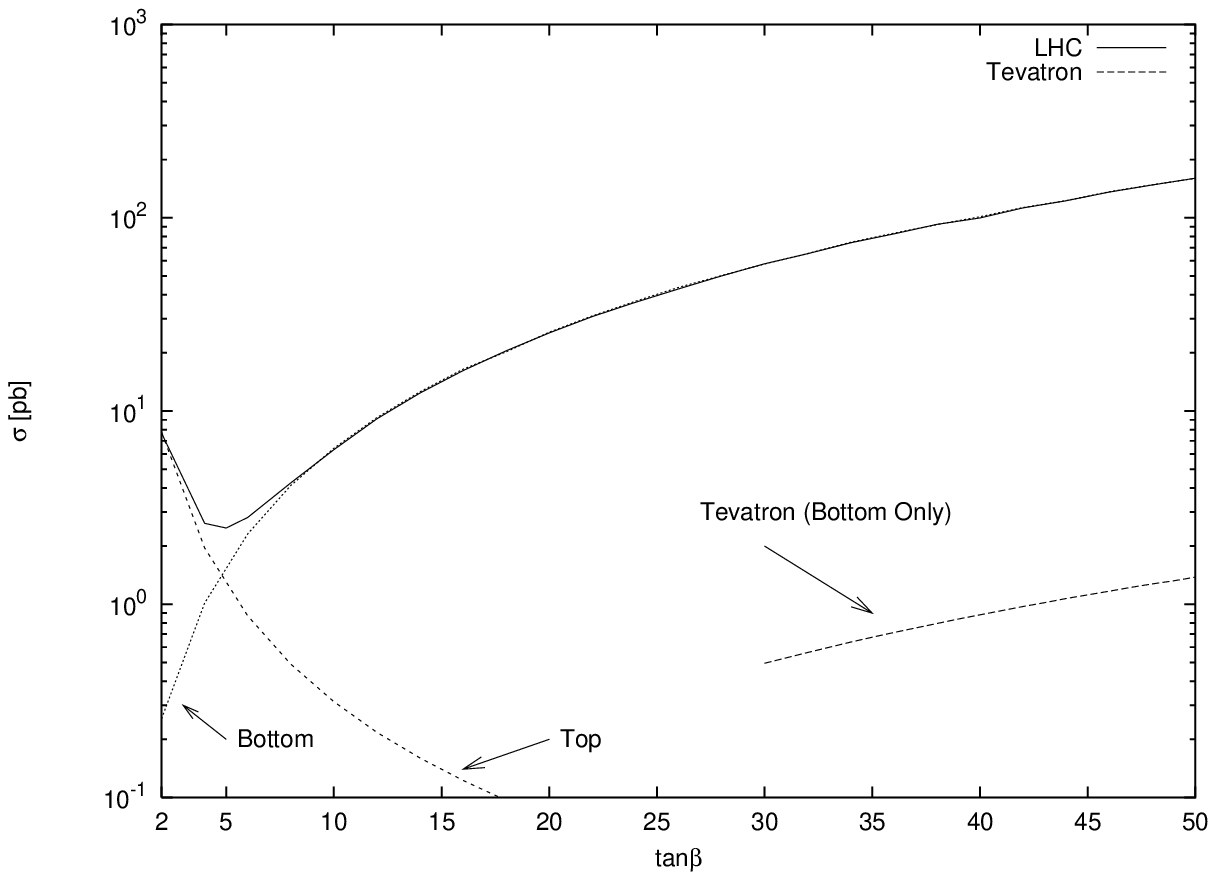}
     \caption{Cross-section for the production of the MSSM
pseudoscalar Higgs boson plus one jet for different values of $\tan\beta$
at the LHC for $M_{A^{0}} = 120$~GeV\!/c$^2$ integrated for
$p_t>30$~GeV\!/c using the full theory matrix elements. The top and bottom
labels show what the contribution of the top and bottom quark would be
alone. In the region $4<\tan\beta<8$ the total cross-section is not
represented well by either the top or bottom matrix elements alone.  In
the experimentally accessible region, the total cross-section at the
Tevatron is dominated by the bottom loop so only the bottom contribution
is shown for $\tan\beta > 30$.}
  \label{total_a_lhc}
\end{figure}

\begin{figure}
  \begin{center}
    \begin{tabular}{c}
      \includegraphics{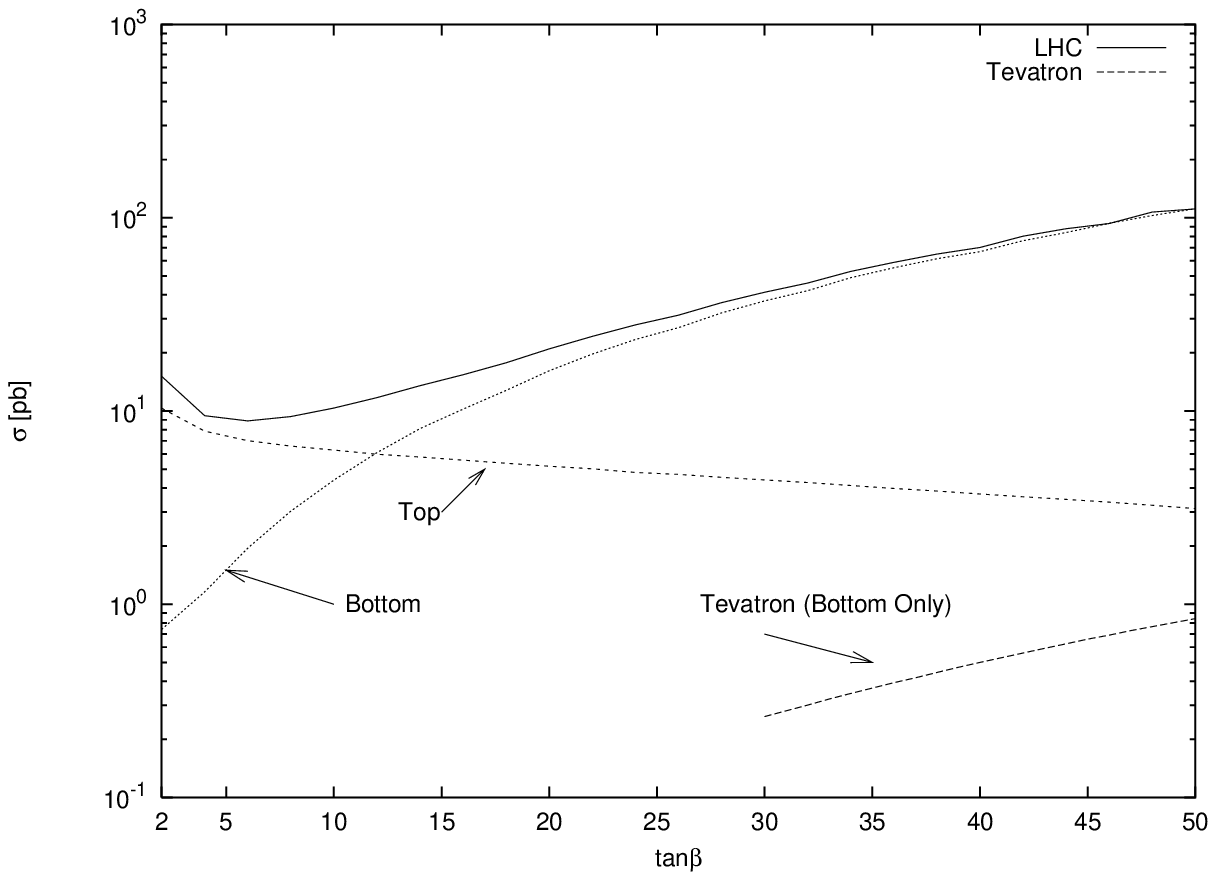} \\
      \includegraphics{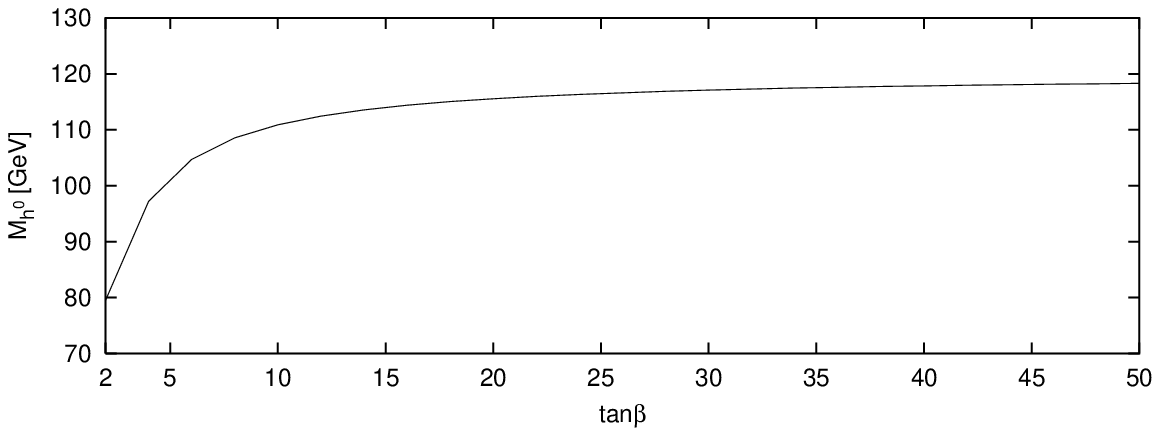}
    \end{tabular}
    \caption{Cross-section for the production of the MSSM lightest
scalar Higgs boson plus one jet for different values of $\tan\beta$
integrated for $p_t>30$~GeV\!/c using the full theory matrix elements. 
The top and bottom labels show what the contributions of the top and
bottom quark would be alone.  In the experimentally accessible region,
the total cross-section at the Tevatron is dominated by the bottom loop,
so only the bottom contribution is shown for $\tan\beta > 30$. Below is
the corresponding mass of the lightest scalar for $M_{A^{0}} =
120$~GeV\!/c$^2$.}
  \label{h_mssm}
  \end{center}
\end{figure}

For the LHC, the entire $\tan\beta$ region is experimentally accessible.  
In the small $\tan\beta$ region, the cross-section is well approximated by
the HQET limit and the bottom quark contribution can be neglected.
However, there are regions where both the top and bottom quark loops are
important. The results are summarized in Figs.~\ref{total_a_lhc} and
~\ref{h_mssm}. These plots use the full theory matrix elements. For
pseudoscalar plus jet production, including only the top quark loop
underestimates the total cross-section by $9.5$\% at $\tan\beta = 4$ and
the discrepancy becomes larger as $\tan\beta$ grows. Including only the
bottom quark underestimates the total cross-section by $5.6$\% at
$\tan\beta=8$ and becomes a better approximation as $\tan\beta$ increases.
The total cross-section for the MSSM lightest scalar plus jet production
receives an important contribution from the interference between the top-
and bottom-quark loops over a large range of $\tan\beta$.

\section{Conclusions}
\label{conclusion}

We calculated the differential distributions and cross-sections for the
SM Higgs, $H$, the MSSM scalar Higgs boson, $h^0$, and pseudoscalar
boson, $A^0$, plus one jet production at the Tevatron and LHC. We
included both the top and bottom quark loops and investigated the
validity of the Heavy Quark Effective Theory (HQET) limit and the light
quark mass limit.  For large $\tan\beta$, the HQET fails and the
complete result with all mass dependences is needed.

The NLO QCD corrections for Higgs plus jet\cite{jack,florian,glosser} and
pseudoscalar plus jet\cite{field2} production have been previously found
in the large $m_\text{top}$ limit. Our results make it clear that these
can only be applied to the MSSM in certain regions. At large $\tan\beta$,
using the bottom-quark only is a very good approximation in the MSSM. At
small $\tan\beta$ the MSSM pseudoscalar is top-quark loop dominated,
whereas the lightest scalar in the MSSM still receives important
contributions from both the top- and bottom-quarks over a much broader
range of $\tan\beta$. This can be seen as the effective suppression of the
$c_t^{h^{0}}$ coupling and enhancement of the $c_b^{h^{0}}$ coupling at
small $\tan\beta$ where the interference between the two terms is still
playing an important role.

\begin{acknowledgments}

B.~Field would like to thank W.~Kilgore for discussions on the
pseudoscalar coupling as well as A.~Field-Pollatou, N.~Christensen and
J.~Ellis for helpful comments and suggestions. The work of B. Field and J.
Smith is supported in part by the National Science Foundation grant
PHY-0098527. The work of S. Dawson is supported by the U.S. Department of
Energy under grant DE-AC02-98CH10886.

\end{acknowledgments}

\appendix

\section{Complete Pseudoscalar Matrix Elements}
\label{pseudo}

For the $q\bar{q} \rightarrow gA^0$ sub-process, the 
(spin and color averaged) matrix elements squared are 
particularly simple because the presence of a $\gamma_5$ makes the traces 
much smaller than in the scalar case. They can be written in terms of the 
integrals presented 
in\cite{baur},
\begin{equation}
\overline{\sum}|\mathcal{M}|^2_{q {\overline q}\rightarrow g A^0}
 = \frac{16 m_f^4}{\hat{s}}
\biggl(\frac{(4\pi\alpha_s(\mu_r^2))^3}{4 N_c^2 v^2}\biggr) \;
|C_1(\hat{s})|^2 \; [ \hat{s}^2 - 2 \hat{t}_1 \hat{u}_1 + Q^4],
\end{equation}
where the new variables are defined
\begin{equation}
      \hat{s}_1 = \hat{s}-Q^2,
\quad \hat{t}_1 = \hat{t}-Q^2,
\quad \hat{u}_1 = \hat{u}-Q^2.
\end{equation}
It is easy to see that $\hat{s}_1 = -(\hat{t}+\hat{u})$ and so on. 

In these expressions we use the notation of \cite{baur}. The $C_1$ loop
integral that appears in the calculation is the usual triangle integral 
with two massive legs. For $p_1^2 =0$, $p_2^2 = Q^2=m_\Phi^2$, $p_{12} = 
p_1+p_2$ and $p_{12}^2=\hat{s}$, the triangle integral is defined as
\begin{align}
C_1(\hat{s}) &= C_1(p_1,p_2) \\ &= \frac{1}{i \pi^2} \int
         \frac{d^4 \! q}
              {[q^2-m_f^2][(q+p_1)^2-m_f^2][(q+p_{12})^2-m_f^2]}.
\end{align}
The box integrals with $p_1^2 = p_2^2 = p_3^3=0$, and 
$p_{123}^2=(p_1+p_2+p_3)^2=
   Q^2$ are defined as
\begin{align}
D(\hat{s},\hat{t}) &= D(p_1,p_2,p_3) \\ &= \frac{1}{i \pi^2} \int
         \frac{d^4 \! q}
              {[q^2-m_f^2][(q+p_1)^2-m_f^2]
                          [(q+p_{12})^2-m_f^2]
                          [(q+p_{123})^2-m_f^2]}.
\end{align}
It is easy to see that the box integrals satisfy the relation 
$D(\hat{x},\hat{y}) = D(\hat{y},\hat{x})$. The 
computer package FF\cite{ff} was used to evaluate 
the scalar integrals.

For the $gg \rightarrow gA^0$
sub-process, the (spin and 
color averaged) matrix element squared can be written in
the symmetric form,
\begin{align}
\overline{\sum}|\mathcal{M}|^2_{gg\rightarrow g A^0}
 = \sum_f\frac{m_f^4(4\pi\alpha_s(\mu_r^2))^3}{v^2(N_c^2-1)^2} \biggl\{ 
  & F(\hat{s},\hat{t},\hat{u})+
    F(\hat{s},\hat{u},\hat{t})+
    F(\hat{u},\hat{s},\hat{t})+
    F(\hat{u},\hat{t},\hat{s})+
    F(\hat{t},\hat{u},\hat{s})+
    F(\hat{t},\hat{s},\hat{u}) \biggr\}
\end{align}
where
\begin{align}
F(\hat{s},\hat{t},\hat{u})=& -2 \; \text{Re} 
           \biggl( C_1(\hat{u})D^*(\hat{u},\hat{s}) \biggr)
           \biggl[ \hat{s}_1 \biggl( \frac{\hat{s} Q^2}{\hat{t}}
                   + \hat{u} \biggr) - \hat{s}\hat{t} - 
                     \frac{\hat{s}^3}{\hat{t}} 
           \biggr]
             - \frac{1}{2} \; \text{Re} 
               \biggl( D(\hat{s},\hat{t})D^*(\hat{u},\hat{s}) \biggr)
               \biggl[ 
                     \hat{t}_1 ( \hat{s}^2 + \hat{s}\hat{t} )
               \biggr] \nonumber \\
           & + 2 \; \text{Re} 
               \biggl( C_1(\hat{t})C_1^*(\hat{u}) \biggr)
               \biggl[ \frac{\hat{t}^2 - \hat{t}_1 Q^2}{\hat{s}} +Q^2
               \biggl( \frac{Q^4 + 2\hat{s}\hat{s}_1}
                            {\hat{u}\hat{t}} \biggr)
             + \hat{t}-3Q^2 + 4\hat{s} \biggr] \nonumber \\
           & - | C_1(\hat{u}) |^2 \biggl[ 
               \frac{Q^6 \hat{u}_1}
                    {\hat{s}\hat{t}\hat{u}}
             + \frac{\hat{s}^2+Q^4}{\hat{t}}
             + \frac{\hat{t}^2+Q^4}{\hat{s}}
             - \frac{4\hat{s}\hat{t}-3Q^4}{\hat{u}}
             - 3 Q^2 
               \biggr] \nonumber \\
           & + \text{Re} \biggl( C_1(\hat{u})D^*(\hat{s},\hat{t}) \biggr)
               \biggl[   \hat{s} \hat{t} 
                       - \hat{s}_1 \hat{t}_1 + Q^2
               \biggl( \frac{\hat{s}_1^2+\hat{s}^2}{\hat{u}} \biggr)
               \biggr] \nonumber \\
           & - \frac{1}{4} |D(\hat{s},\hat{t})|^2
               \biggl(
               2 (\hat{s}^3+\hat{t}^3) 
               - \frac{\hat{s}^2\hat{t}^2}{\hat{u}}
               + \frac{\hat{s}\hat{t}}{\hat{u}^2}
               \biggl[
               (\hat{s}+\hat{t})^3 - 2 \hat{s}\hat{t}\hat{u}
               \biggr]
               \biggr).
\end{align}

\section{Analytic Limits of Matrix Elements}
\label{analytic}

The partonic  cross-section for $q \overline{q} \rightarrow g \Phi$  is
\begin{equation}
\frac{d\hat{\sigma}}{d\hat{t}} = \frac{1}{16\pi \hat{s}^2} 
                                 \frac{1}{36}
|\mathcal{M}|^2_{q {\overline q}\rightarrow g \Phi},
\label{qqdsig}
\end{equation}
where the spin and color average is explicitly given,
\begin{equation}
|\overline{\mathcal{M}}|^2_{q\overline{q}\rightarrow g \Phi}
\equiv
 \frac{1}{36}|\mathcal{M}|^2_{q {\overline q}\rightarrow g \Phi}.
\end{equation}
For a scalar Higgs,
\begin{equation}
|\mathcal{M}|^2_{q{\overline{q}}\rightarrow g H} 
   = \frac{16 \alpha_s^3}{\pi v^2}
        \frac{\hat{t}^2+\hat{u}^2}{\hat{s}}
        | A_{q \overline{q}}^H|^2,
\end{equation}
and 
\begin{align}
A_{q\overline{q}}^H  = \sum_j 
                       \biggl\{ \frac{m_j^2}{\hat{s}-M_{H}^2}
                     & \biggl[2-\frac{2\hat{s}}{\hat{s}-M_{H}^2}
                       \biggl( I_1(\hat{s} / m_j^2)
                              -I_1(M_{H}^2 / m_j^2)
                       \biggr) \nonumber \\
                    &+ \biggl( 1+\frac{4m_j^2}{\hat{s}-M_{H}^2}
                       \biggr)
                       \biggr( I_2(\hat{s} / m_j^2)
                              -I_2(M_{H}^2 / m_j^2)
                       \biggr)
                       \biggr]
                       \biggr\},
\end{align}
where $m_j$ is the fermion mass in the loop. The integrals are defined by:
\begin{equation}
I_1(a) =\int_0^1 dx \log\biggl(1-ax(1-x)\biggr), \quad
I_2(a) = \int_0^1 \frac{dx}{x}\log\biggl(1-ax(1-x)\biggr).
\end{equation}
In the large fermion mass limit, $m_j\rightarrow \infty$,\cite{baur,ellis}
\begin{equation}
A_{q \overline{q}}^H \rightarrow
  -\frac{1}{3}
   \biggl(1+\frac{1}{120}
            \frac{11\hat{s}+7M_{H}^2}{m_j^2}+\ldots\biggr).
\end{equation}
In the small fermion mass limit, $m_j\rightarrow 0$,\cite{baur}
\begin{align}
A_{q \overline{q}}^H  &\rightarrow 
                            A_{q \overline{q}}^{Hr}
                        + i A_{q \overline{q}}^{Hi}, \nonumber \\
A_{q \overline{q}}^{Hr} & \rightarrow 
                          \frac{2m_j^2}{\hat{s}-M_H^2} 
                          \biggl\{ 1+\Lambda_s 
                          \biggl( -\frac{\hat{s}}{\hat{s}-M_H^2}
                         +\frac{1}{4} 
                          \biggl(
                        1+\frac{4m_j^2}{\hat{s}-M_{H}^2}
                          \biggr)
                          \biggl[ \Lambda_s-2 \log
                          \biggl( \frac{m_j^2}{M_{H}^2}
                          \biggr)
                          \biggr]
                          \biggr)
                          \biggr\}, \nonumber \\
A_{q \overline{q}}^{Hi} & \rightarrow 
                         -\frac{m_j^2\pi}{\hat{s}-M_{H}^2}
                          \biggl( 1+\frac{4m_j^2}{\hat{s}-M_{H}^2}
                          \biggr) \Lambda_s,
\end{align}
where $\Lambda_s=\log(\hat{s} / M_{H}^2)$.

The result for $q g\rightarrow q \Phi$ can be found from crossing,
\begin{equation}
\frac{d\hat{\sigma}}{d\hat{t}} = \frac{1}{16\pi \hat{s}^2}
                     \biggl( \frac{1}{96}\biggr)| M|^2_{q g\rightarrow q \Phi},
\end{equation}
and
\begin{equation}
 |\mathcal{M}(\hat{s},\hat{t},\hat{u}) |^2_{q g\rightarrow q \Phi}=
-|\mathcal{M}(\hat{u},\hat{t},\hat{s}) |^2_{q {\overline q}\rightarrow g \Phi}.
\end{equation}
In the large fermion mass limit, $m_j\rightarrow \infty$\cite{baur,ellis},
\begin{equation}
A_{q g}^H   \rightarrow
           -\frac{1}{3}
            \biggl( 1+\frac{1}{120}
                      \frac{11\hat{u}+7M_{H}^2}{m_j^2}+ \ldots
            \biggr)
\end{equation}
In the small fermion mass limit, $m_j\rightarrow 0$,
\begin{align}
A_{q g}^H & \rightarrow   A_{q g}^{Hr} +
                          i A_{q g}^{Hi} \nonumber, \\
A_{q g}^{Hr}& \rightarrow  \frac{2m_j^2}{\hat{u}-M_{H}^2}
                           \biggl\{ 1+\Lambda_u
                           \biggl(-\frac{\hat{u}}{\hat{u}-M_{H}^2}
                          +\frac{1}{4}
                           \biggl(1+\frac{4m_j^2}{\hat{u}-M_{H}^2}
                           \biggr)
                           \biggl[ \Lambda_u-2 \log
                           \biggl( \frac{m_j^2}{M_{H}^2}
                           \biggr)
                           \biggr]
                           \biggr)
                           \biggr\}, \nonumber \\
A_{q g}^{Hi}& \rightarrow -\frac{2 m_j^2 \hat{u}\pi}{(\hat{u}-M_{H}^2)^2},
\end{align}
where $\Lambda_u=\log(|\hat{u}| / M_{H}^2)$.

The results for pseudoscalar production are found assuming the
$\bar{\psi}\psi A^0$ coupling given in Eq. \ref{pseudodef}. The
form factor for
\begin{equation}
g(p_1^{\mu_1,a_1})+g(p_2^{\mu_2,a_2}) \rightarrow A^0(p_5),
\end{equation}
with all moment outgoing and $p_1^2=0$, $p_5^2=M_{A^0}^2$, 
$(p_1+p_5)^2=\hat{s}$, is
given by
\begin{align}
i\Gamma^{\mu_1,\mu_2}(p_1,p_2,p_5) &= -\frac{\alpha_s}{2 \pi}
                                   \frac{g_A m_j^2}{v}
                                   \delta_{a_1 a_2}
                                   \epsilon^{\alpha\beta\mu_1\mu_2}
                                   p_5^{\alpha} p_2^{\beta} \nonumber \\
                               &   \qquad \times
                                   \frac{1}{\hat{s}-M_{A^{0}}^2}
                                   \biggl\{ I_2(\hat{s} / m_j^2)
                                           -I_2(M_{A^{0}}^2 / m_j^2)
                                   \biggr\}.
\end{align}

The differential cross-section for $q \overline{q}\rightarrow g A^0$
is given by Eq. \ref{qqdsig}, with
\begin{equation}
\label{qqgA}
| \mathcal{M} |^2_{q {\overline q}\rightarrow g A^0}=
  \frac{\alpha_s^3}{\pi} g_A^2 \sum_j
  \frac{4 m_j^4}{\hat{s} v^2}
  \biggl( 1+\frac{2\hat{t}\hat{u}}{(\hat{s}-M_{A^{0}}^2)^2}
  \biggr) \biggl| I_2(\hat{s} / m_j^2)
                 -I_2(M_{A^{0}}^2 / m_j^2) \biggr|^2.
\end{equation}
In the large fermion mass limit, $m_j\rightarrow \infty$\cite{kao},
\begin{equation}
| \mathcal{M}|^2_{q {\overline q}\rightarrow g A^0} \rightarrow 
  \frac{\alpha_s^3}{\pi} g_A^2 
  \frac{1}{\hat{s} v^2}
  \biggl( 2\hat{t}\hat{u} + (\hat{s}-M_{A^{0}}^2)^2
  \biggr)
  \biggl[1+ \frac{\hat{s}+M_{A^{0}}^2}{6 m_j^2}+ \ldots
  \biggr].
\end{equation}
In the small fermion mass limit, $m_j\rightarrow 0$,
\begin{equation}
| \mathcal{M}|^2_{q {\overline q}\rightarrow g A^0} \rightarrow 
  \frac{\alpha_s^3}{\pi} g_A^2
  \frac{1}{\hat{s} v^2}
  \biggl( 1+\frac{2\hat{t}\hat{u}}{(\hat{s}-M_{A^{0}}^2)^2}
  \biggr) m_j^4 \Lambda_{s}^2
  \biggl\{
  \biggl[ \Lambda_s-2 \log
  \biggl( \frac{m_j^2}{M_{A^{0}}^2}
  \biggr)
  \biggr]^2 + 4\pi^2
  \biggr\},
\end{equation}
where $\Lambda_s=\log(\hat{s} / M_{A^{0}}^2)$.


\begin{thebibliography}{2003}
\bibitem{footnote}
There is some confusion over the coupling constant for the pseudoscalar
case in the literature. The correct coupling is found in
Ref.~\cite{schaffer}. There is an extra factor of $1/4$ in
Ref.~\cite{russel2} leading to a cross-section $16$ times too small for
the pseudoscalar case. It seems that the $1/4$ from the effective
Lagrangian was incorporated into the coupling constant by mistake. The
Feynman rules in both papers are correct if the coupling constant from
the Ref.~\cite{schaffer} paper is used.
\end{thebibliography}
\end{document}